\documentclass[a4paper,12pt]{article}
\usepackage[utf8]{inputenc}
\usepackage{amsmath}
\usepackage{t1enc}
\usepackage{anysize}
\usepackage{amsfonts}
\usepackage{amssymb}
\usepackage{graphicx}
\marginsize{1.5cm}{1.5cm}{1.6cm}{1.6cm}
\title{Virus spread and voter model on random graphs with multiple type nodes}

\newtheorem{Theorem}{Theorem}
\newtheorem{definition}{Definition}

\usepackage{graphicx}

\begin{document}

\begin{center}
{\LARGE\bf \textsc{Virus spread and voter model on random graphs with multiple type nodes}}
\vspace*{1cm}
\\
{\large Ágnes Backhausz}\footnote{ELTE Eötvös Loránd University, Budapest, Hungary, Faculty of Science, Department of Probability and Statistics and Alfréd Rényi Institute of Mathematics. }

{\large Edit Bognár}\footnote{ELTE Eötvös Loránd University, Budapest, Hungary, Faculty of Science.}
 
\end{center}

\begin{abstract}
    When modelling epidemics or spread of information on online social networks, it is crucial to include not just the density of the connections through which infections can be transmitted, but also the variability of susceptibility. Different people have different chance to be infected by a disease (due to age or general health conditions), or, in case of opinions, ones are easier to be convinced by others, or stronger at sharing their opinions. The goal of this work is to examine the effect of multiple types of nodes on various random graphs such as Erdős--Rényi random graphs, preferential attachment random graphs and geometric random graphs. We used two models for the dynamics: SEIR model with vaccination and a version of voter model for exchanging opinions.  In the first case, among others, various vaccination strategies are compared to each other, while in the second case we studied sevaral initial configurations to find the key positions where the most effective nodes should be placed to disseminate opinions.      
\end{abstract}

\vspace*{0.3cm}

{\bf Acknowledgement.} The project was  supported by the European Union, co-financed by the European Social Fund (EFOP-3.6.3-VEKOP-16-2017-00002).

\section{Introduction}
Creating mathematical models for interacting particle systems was motivated by numerous real life processes from the field of biology and physics. In a network, particles can have different characteristic features of many aspects, which could affect the dynamics of stochastic processes performed on them. 
In this paper we examine two similar, yet different stochastic processes, where individuals spread opinions or a virus between one another depending on relationships or social contacts between them. The common feature in our two models is that we model the network by random graphs where nodes have different types, that is, their susceptibility for interactions with other vertices is different. In addition, in many cases, this type is chosen randomly, in a way which is closely related to the structure of the dynamically evolving network. On the other hand, we also ask what is the optimal way of choosing the types when the total number of vertices of a type is given, but there is some freedom in choosing the position of these vertices.
\\
One of our main interests is epidemic spread. The accurate modelling, regulating or preventing of a possible epidemic is still a difficult problem of the 21st century. (As of the time of writing, a novel strain of coronavirus has spread to at least 16 other countries from China, although authorities have been taking serious actions to prevent a worldwide outbreak.) As for mathematical modelling, there are several approaches to model these processes, for example, using differential equations, the theory of random graphs or other probabilistic tools \cite{durrett2, hofstad, simon}. As it is widely studied,  the structure of the underlying graph can have an important impact on the course of the epidemic. In particular, structural  properties such as degree distribution and clustering are essential to understand the dynamics and to find the optimal vaccination strategies \cite{britton2, fransson}. From the point of view of random graphs, in case of preferential attachment graphs, it is also known the initial set of infected vertices can have a huge impact on the outcome of the process \cite{berger}: A small proportion infected vertices is enough for a large outbreak if the positions are chosen appropriately. On the other hand, varying susceptibility of vertices also has an impact for example on the minimal proportion of vaccinated people to prevent the outbreak \cite{britton, bhansali}. In the current work, by computer simulations, we study various cases when these effects are combined in a SEIR model with vaccination: We have a multitype random graph, and the  vaccination strategies may depend on the structure of the graph and types of the vertices as well.   
\\
The other family of models which we studied is a variant of the voter model. The voter model is also a common model of interacting particle systems and population dynamics, see e.g.\ the book of Liggett \cite{3}. This model is related to epidemics as well: Durett and Neuhauser \cite{durrett} applied the voter model to study virus spread. The two processes can be connected by the following idea:  We can see virus spread as a special case of the voter model with two different opinions (healthy and infected), but only one of the opinions (infected) can be transmitted, while any individuals with infected opinion switch to healthy opinion after a period of time. Also the virus can spread only through direct contacts of individuals (edges of the graphs), while in the voter model it is possible for the particles to influence one another without being neighbors in the graph. Similarly to the case of epidemics, the structure of the underlying graph has an important impact on the dynamics of the process \cite{basu, carro}. Here we study a version of this model with various underlying random graphs and multiple types of nodes. 
\\
We examined the virus spread with vaccination and the voter model on random graphs of different structures, where in some cases the nodes of the graph corresponding to the individuals of the network are divided into groups representing significantly distinct properties for the process.
We studied the possible differences of the processes on different graphs, regarding the nature and magnitude of distinct result and both tried to find the reasons for them, to understand how can the structure of an underlying network affect outcomes.
\\
The outline of the paper is as follows. In the second section we give a description of the virus spread in continuous time, and the discretized model. Parameters are chosen such that they match the real-world data from \cite{1}. We confront outcomes on different random graphs and the numerical solutions of the differential equations originating from the continuous time counterpart of the process. We also study different possible choices of reproduction number $R_0$ corresponding to the seriousness of the disease. We examine different vaccination strategies (beginning at the start of the disease of a few days before), and a model with weight on edges is also mentioned.
\\
In the third section we study the discretized voter model on Erdős--Rényi and Barabási--Albert graphs firstly without, then with multiple type nodes. Later we run the process on random graphs with a geometric structure on the plane.

\section{Virus spread}

The dynamics of virus spread can be described by differential equations, therefore they are usually studied from this approach. However, differential equations use only transmission rates calculated by the number of contacts in the underlying network, while the structure of the whole graph and other properties are not taken into account. Motivated by the paper   "Modelling the strategies for age specific vaccination scheduling during influenza pandemic outbreaks" of Diána H. Knipl and Gergely Röst \cite{1}, we modelled the process on random graphs of different kinds. In this section we use the same notions and sets for most of the parameters. Ideas for vaccination strategies are also derived from there.

\subsection{The model}

We examined a model in which individuals experience an incubation period, delaying the process. Dynamics are also effected by a vaccination campaign started at the outbreak of the virus, or vaccination campaign coming a few days before the outbreak. 

\begin{definition}
In the \emph{classical SEIR model}, each individual in the model is in exactly one of the following compartment during the virus spread:
\begin{itemize}
    \item Susceptible: Individuals are healthy, but can be infected.
    \item Exposed: Individuals are infected but not yet infectious.
    \item Infectious: Individuals are infected and infectious.
    \item Recovered: Individuals are not infectious anymore, and immune (cannot be infected again).
\end{itemize}

Individuals can move through compartments only in the defined way above (it is not possible to miss out one in the line). The rate at which individuals leave compartments are described by probabilities (transmission rates) and the parameters of the model (incubation rate, recovery rate).  Individuals in $R$ are immune, so $R$ is a terminal point.

\end{definition}

SEIR with vaccination:
We mix the model with a vaccination campaign. The campaign lasts for 90 days, and we vaccinate individuals according to some strategy (described later) so that at the end of the campaign $60\% $ of the population is vaccinated (if it is possible). We vaccinate individuals only in $S$, but the vaccination ensures immunity only with probability $q$, and only after 14 days. We vaccinate individuals at  most once irrespectively of the success of vaccination. However, vaccinated individuals can be infected within the first 14 days. In this case, nothing differs from the process without vaccination.

\begin{figure}
    \centering
    \includegraphics[height=5cm]{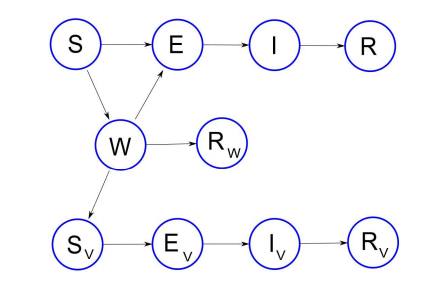}
    \caption{Transmitting stages in SEIR with vaccination [1]}
    \label{fig:seir}
\end{figure}

Figure \ref{fig:seir} from [1] summarizes the stages of the whole process. $W$ denotes vaccinated individuals, $R_W$ individuals immune to the infection due to vaccination, and $V$-indexed stages correspond to unsuccessful vaccinations. (Individuals can leave a stage only on an edge heading out.)

\subsection{Data}
To describe the underlying network, we use real-life data. We distinguish individuals according to their age. In particular, we consider 5 age groups since they have different social contact profile. The age groups and the number of individuals in each per $N=10000$ citizen (distributed as the 2005 European Union population, Eurostat 2006):
\begin{itemize}
    \item 0-9: $N^1=1050$
    \item 10-19: $N^2=1200$
    \item 20-39: $N^3=2850$
    \item 40-65: $N^4=3250$
    \item 65+: $N^5=1650$
\end{itemize}
To describe the social relationships of the different age groups, we used the contact matrix obtained in [1]:
\[
C=
\begin{pmatrix}
5,3580 & 1,0865 & 3,0404 & 2,4847 & 0,8150 \\
0,9507 & 10,2827 & 2,8148 & 3,6215 & 0,7752 \\
1,1201 & 1,1852 & 6,5220 & 4,1938 & 0,9016 \\
0,8027 & 1,3372 & 3,6776 & 5,2632 & 1,3977 \\
0,5187 & 0,5638 & 1,5573 & 2,7531 & 2,0742 \\
\end{pmatrix},
\] where the elements $c_{i,j}$ represent the average number of contacts an individual in age group  $i$ has with individuals in age group $j$.

\subsection{Parameters}
\label{parameters}
In the sequel, the number of individuals in a given group is denoted by the label of the group according to Figure \ref{fig:seir}.
The model is specified by the following family of parameters.
\begin{itemize}
\item $R_0=1.4$: basic reproduction number. It characterizes the intensity of the epidemic. Its value is the average number of infections an infectious individual causes during its infectious period in a population of only susceptible individuals (without vaccination). Later we also study less severe cases with $R_0=1.0-1.4$.
\item$\beta_{i,j}$: transmission rates. They control the rate of the infection between a susceptible individual in age group $i$ and an infectious individual in age group $j$. They can be derived from $R_0$ and the $C$ contact matrix. 
According to [1] we used $\beta_{i,j} = \beta \cdot \frac{c_{i,j}}{N^j}$, where $\beta= 0.0334$ for $R_0=1.4$.
\item$\frac{1}{\nu_E}=\frac{1}{\nu_{E_V}}=1.25$: latent period. $\nu_E$ is the rate of exposed individuals becoming infectious.
\item$\frac{1}{\nu_{I}}=\frac{1}{\nu_{I_V}}=3$: infectious period. Each individual spends an average of $\frac{1}{\nu_I}$ in $I$.
\item$\frac{1}{\nu_W}=14$: time to develop antibodies after vaccination.
\item$q_i=0.8$ for $i=1, \dots, 4$ and $q_5=0.6$:  vaccine efficacy. The probability that a vaccinated individual develops antibodies and becomes immune.
\item$\delta=0.75$: reduction in infectiousness. The rate by which infectiousness of unsuccessfully vaccinated individuals is reduced.
\item$\lambda^{i}= \sum_{j=1}^5 \beta_{j,i} \cdot(I^{j}+\delta \cdot I^{j}_V)$ is the total rate at which individuals of group $i$ get infected and become exposed.
\item $V^{i}$: vaccination rate functions determined by a strategy. This describes the rate of vaccination in group $i$.
\end{itemize}

The dynamics of the virus spread and the vaccination campaign can be described by 50 differential equations (10 for each age group), according to \cite{1}:

\begin{tabular}{l l l l l}
$(S^{i})'=-S^{i}\cdot \lambda^{i}-V^{i}$ & & & & $
(S^{i}_V)'=(1-q_i)\nu_W\cdot W^{i}-S_V^{i} \cdot \lambda^{i}$\\
$
(E^{i})'=(S^{i}+W^{i})\cdot \lambda^{i}-\nu_E \cdot E^{i}
$ & & & &$
(E^{i}_V)'=S^{i}_V \cdot \lambda^{i}-\nu_{E_V}\cdot E_V^{i}$\\
$
(I^{i})'=\nu_E \cdot E^{i}-\nu_I \cdot I^{i}
$ & & & & $
(I^{i}_V)'=\nu_{E_V}\cdot E_V^{i}- nu_{I_V}\cdot I_V^{i}$ \\
$
(R^{i})'=\nu_I \cdot I^{i}
$ & & & & $
(R^{i}_V)'=nu_{I_V}\cdot I_V^{i}
$ \\
$
(W^{i})'=V^{i}-W^{i}\cdot \lambda^{i} -\nu_W \cdot W^{i}
$ & & & & $
(R^{i}_W)'=q_i\cdot \nu_{W}\cdot W^{i}
$ \\
\end{tabular}
\\

\subsection{Random graphs}
We would like to create an underlying network and examine the outcome of virus spread on this given graph. 

We generated random graphs of different structures with $N=10000$ nodes, such that each node has a type corresponding to the age of the individual. The age distributions and number of contacts in the graph between age groups comply with statistic properties detailed above. Since the contact matrix $C$ describes only the average number of contacts, the variances can be different. 
\begin{itemize}
    \item Erdős--Rényi graphs: We create $10000$ nodes and their types are defined immediately, such that the number of types comply exactly to the  age distribution numbers. The relationships within each age group and the connections between different age groups are both modelled with an Erdős--Rényi graph in the following sense: We create an edge between every node in age group $i$ and node in age group $j$ independently with probability $p_{i,j}$, where $p_{i,j}=\frac{c_{i,j}}{N^j}$ if $i\neq j$, and $p_{i,j}=\frac{c_{i,j}}{N^j-1}$ if $i=j$.
    
    \item Preferential attachment graphs: Initially we start from an Erdős--Rényi graph of size 100, then we keep adding nodes to the graph sequentially. Every new node chooses its type randomly, with probabilities given by the observed age distribution. After that we create edges between the new node and the old ones with preferential attachment. If the new node is of type $i$, then  we connect it with an edge independently to an old node $v$ of type $j$ with probability $\frac{c_{i,j}\cdot d(v)}{2 \cdot D}$, where $d(v)$ denotes the actual degree of $v$, and $D$ is the sum of degrees belonging to nodes with type $j$. Thus the new node is more likely to attach to nodes with a high degree, resulting a few enormous degrees in each age group. On the other hand, the connection matrix $C$ is used to ensure that the density of edges between different age groups is different.
    
    \item Preferential attachment mixed with Erdős--Rényi graphs: We create the $10000$ nodes again with their types exactly according to age distribution numbers. 
    First we create five preferential attachment graphs, the $i$th of size $N^i$ so that every node has an average of $c_{i,i}$ neighbours. In particular, the endpoints of the new edges are chosen independently, and the attachment probabilities are proportional to the degrees of the old vertices. Then we attach nodes in different age groups independently with the corresponding $p_{i,j}$ probabilities defined above.
    
    \item Random graphs of minimal degree variances with configuration model: We prescribe not only a degree sequence of the nodes, but the degree of a node broken down into $5$ parts regarding the age groups in a way the expectations comply with the contact matrix $C$, but the degrees also have a small variance. The distribution is chosen such that the variance is minimal among distribution supported on the integers, given the expectation. For example in case of $c_{1,4}=2,4847$ every node in age group 1 has exactly 2 or 3 neighbours in age group 4, and the average number is $2,4847$. 
    Our configuration model creates a random graph with the given degree sequence.  According to \cite{4}, the expected value of the number of loops and multiple edges divided by the number of nodes tends to zero, thus for $N=10000$ it is suitable to neglect them and to  represent a network of social contacts with this model.
\end{itemize}

\subsection{Discretized SEIR model with vaccination for graphs}

In this section we detail how we implemented the discretization of the process on the generated random graphs. Most of the parameters remained the same as in the differential equations, however to add more reality we toned them with a little variance. For the matter of the transmission rates, we needed to find different numbers to describe the probability of infections, since $\beta$ was derived from $C$ matrix and $R_0$ basic reproduction number. Since $C$ is built in the structure of our graphs,  using different parameters $\beta_{i,j}$ would result in adding the same effect of contacts twice to the process. Therefore instead of $\beta_{i,j}$, we determined a universal $\tilde\beta$ according to the definition of $R_0$.  We set the disease transmissions to $\tilde\beta=\frac{R_0}{3\cdot 12.8113}$, under the assumption, that the contact profile of age groups are totally implemented in the graph structure. Only the average density of the graph (without age groups), severity of the disease and average time spent in infectious period can affect the parameters.
Parameters $\nu_W$, $\delta$, $q_i$ remained exactly the same, while $\frac{1}{\nu_E}=1.25$ and $\frac{1}{\nu_{I}}=3$ holds only in expected value. The exact distributions are given as follows: 
\[\mathbb{P}\bigg(\frac{1}{\nu_{I}}=2\bigg)=\mathbb{P}\bigg(\frac{1}{\nu_{I}}=4\bigg)=\frac{1}{4}, \qquad \mathbb{P}\bigg(\frac{1}{\nu_{I}}=3\bigg)= \frac{1}{2}\] and \[\mathbb{P}\bigg(\frac{1}{\nu_{E}}=1\bigg)=0.75, \qquad \mathbb{P}\bigg(\frac{1}{\nu_{E}}=2\bigg)=0.25.\]
\\ We built in reduction in infectiousness $\delta$ in the process in such way that an unsuccessfully vaccinated individual spends $3\cdot 0.75=2.25$ days in average in $I$, instead of modifying $\tilde\beta$.

In the discretized process, we start with 10 infectious nodes chosen randomly and independently from the age groups. We observe a 90 day period with vaccination plus 10 days without it. (In the basic scenarios we start vaccination at day 1, however we later examine the process with vaccination starting a few days before the outbreak). At a time step firstly the infectious nodes can transmit the disease to their neighbours. Only nodes in $S$ can be infected, and they cannot be infected ever again. When a node becomes infected, its position is set immediately to $E$, and also the number of days it spends in $E$ is generated. Secondly we check if a node reached the end of its latent/infectious period, and we set its position to $I$ or $R$. (As soon as a node becomes infectious, the days it spends in $I$ is also calculated.) Then at the end of each iteration we vaccinate $0.67\%$ of the whole population according to some strategy (if it is possible). Only nodes in $S$ get vaccination (at most once), it is generated immediately whether the vaccination is successful (with probability $q_i$, according to its type). In case of success, the day it could become immune without any infection is also noted. If it reaches the 14th day, and still in $S$, its position is set to immune.

\subsection{Results on the basic scenario}

The first question is whether the structure of the underlying graph can affect the process, in the case when the edge densities are described by the same contact matrix $C$. We can ask how it  can affect the overall outcome and other properties, and how  we can explain and interpret these differences regarding the structure of the graph. We compare results on different graphs with each other, and also with the numerical solution of the differential equations describing the process.
In this section we study the basic scenario: Our vaccination starts at day 1, we vaccinate by uniform strategy. This strategy does not distinguish age groups, every day we vaccinate  $0.67\%$ of each age group randomly (if it is still possible). We set $R_0=1.4$.

Using the differential equation system from \cite{1} detailed in \ref{parameters} sequel, we calculated the numerical solution with $RK4$ method in Matlab, using $\beta_{i,j}$ parameters (concerning still $R_0=1.4$). The solution starts from $[1.05, 1.20, 2.85, 3.25,1.65]$ infected individuals in age groups, since this is the average value of initial infectious cases in our simulated graphs. (Initially we have no individuals in $E, R$ and $W$.)
In the numerical solution 13.562 $\%$ of the population was affected by the disease at the end of the 100 days. Infections reached its peak around day 45-50 (different age groups can peak at different time). Age group 2 reaches the peak significantly sooner than the others; age groups 1-5 and 3-4 have similar properties both in magnitude and timing. Approximately maximum 84 individuals were in $I$ stage at the same time.

\begin{figure}[h!]
    \centering
    \includegraphics[height=3.5cm]{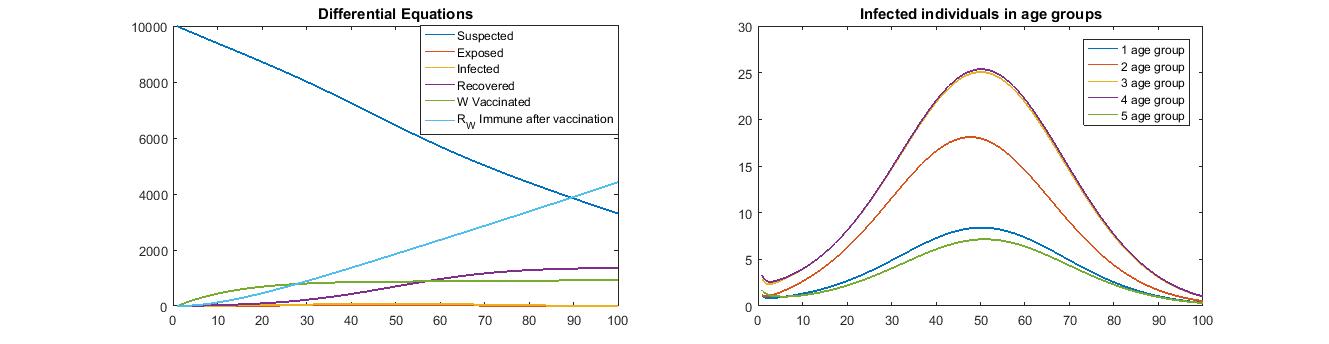}
    \caption{Numerical solution of differential equations}
    \label{fig:diffegyek}
\end{figure}

Giving structure to the underlying social network boosted these numbers in every case, however differences are still significant between the random graphs of different properties.

To study the result on the discretized model, we generated 5 random graphs with $N=10000$ nodes for each graph structure, and run the process 20 times on a random graph with independent initial choice of infected individuals. All in all, these 100 results were averaged. (Creating these graphs with $10000$ nodes is rather time-consuming, while running the virus spread process is relatively fast. In case of most of the structures we can derive rather different outcomes on the same graph with different initial values concerning the peak of the virus. Therefore using the same graphs more is acceptable.)

\begin{figure}
    \centering
    \includegraphics[height=7cm]{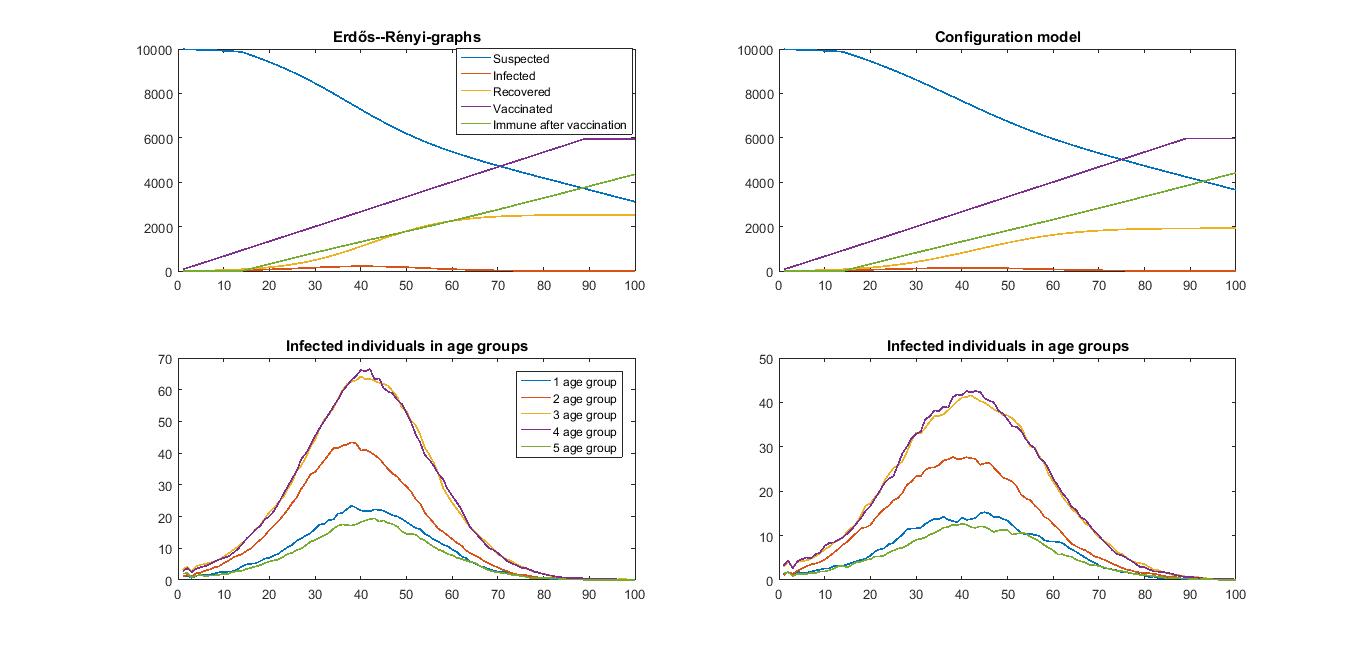}
    \includegraphics[height=7cm]{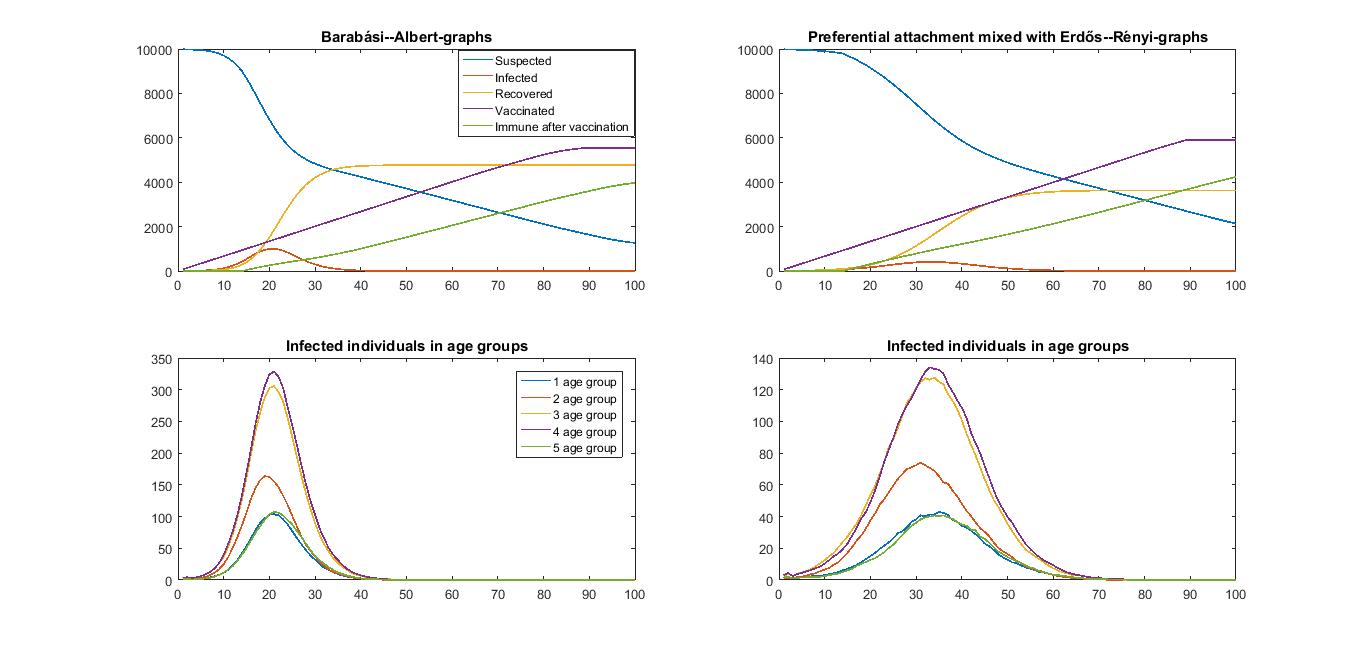}
    \caption{Dynamic of SEIR model and infected individuals in age groups in different random graphs}
    \label{fig:basic}
\end{figure}

As we can see on Figure \ref{fig:basic} (compared to Figure \ref{fig:diffegyek}), random graphs from the configuration model were the closest to the numerical solution of differential equations. However, the difference in outcomes can be clearly seen from every perspective: Almost $20\%$ of the population ($5.7\%$ more) was infected by the virus at the end of the time period, the infection peaked almost $10$ days sooner (at day $41$) and the number of infectious cases at the peak is almost twice as large. We got similar, but more severe result on Erdős--Rényi graphs. however still only a maximum  $0.021\%$  of the population was infected at the same time.
\\
The outcome in case of graphs with a (partial) preferential attachment structure shows that distribution of degrees do matter in this process. (This notice gave the idea initially to model a graph with minimal degree deviation with the help of the configuration model. We were curious if  we can get closer results to the differential equations on such a graph.) On preferential attachment graphs $47.66\%$ of the individuals came through the disease. What is more, $1\%$ of the population was infected at the same time at the peak of the virus, only at day $21$. However, after day $40$ the infection was substantially over. With preferential attachment structure it is very likely that a node with huge degree gets infected in the early days of the process, irrespectively of initial infectious individual choices, resulting in an epidemic really fast. However, after the dense part of the graph passed through the virus around day $40$, even $40\%$ of the population is still in $S$,  magnitude of infectious cases is really low.
\\
The process on Preferential attachment mixed with Erdős--Rényi graphs reflects something in between, yet preferential properties dominate.
It was possible to reach $60\%$ vaccination rate during the process, except in case of Preferential attachment graphs. At the end of the 100th day, $0.4-0.45$ proportion of individuals could acquire immunity after vaccination. 

\begin{center}
\begin{tabular}{ |p{3cm}||p{3cm}|p{3cm}|p{2cm}|  }
 \hline
 \multicolumn{4}{|c|}{Properties} \\
 \hline
 Random graph structure & Overall attack rate & Maximum number of infection at peak & Peak day\\
 \hline
 Differential equations   & 0.1356   & 84 &   50\\
 Erdős--Rényi-graphs &   0.2526 & 211.4 & 40 \\
 Configuration Model & 0.1923 & 137.38 &  41\\
 Preferential attachment  & 0.4766 & 1000&  21\\
 Preferential a. mixed with ER & 0.3632 & 412.44  & 33\\
 \hline
\end{tabular}
\end{center}

\subsection{Sensitivity to changes in parameters}

Basic reproduction number is a representative measure for the seriousness of disease. Generally, diseases with a reproduction number greater than 1 should be taken seriously, however the number is a measure of potential transmissibility. It does not actually tell, how fast a disease will spread. Seasonal flu has an $R_0$ about $1.3$, HIV and SARS around $2-5$, while according to \cite{5} $R_0$ of the coronavirus (2019-nCoV) pneumonia outbreak ranges from 3.30 to 5.47, and significantly larger than 1 in the early phase of the outbreak.\\

\begin{figure}[h!]
    \centering
    \includegraphics[height=4cm]{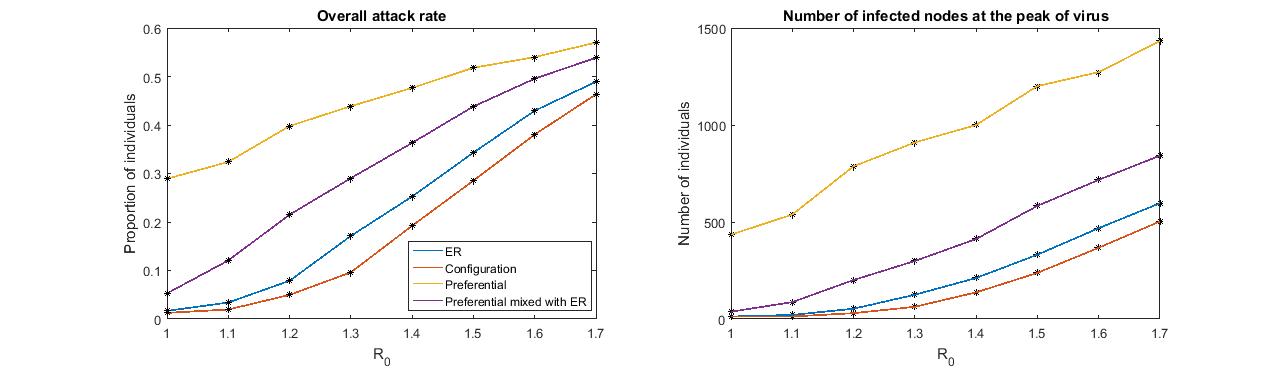}
    \caption{Overall attack rates and maximum number of infected individuals for different $R_0$ reproduction numbers}
    \label{fig:sensitivity}
\end{figure}

In this section we investigate how can different $R_0=[1.0, 1.7]$ parameters affect the outcome of the process comparing different structures of random graphs (see Figure \ref{fig:sensitivity}. The severity of the epidemic for graph structures, both from the aspect of overall attack rate, and number of infected individuals at the peak of the epidemic, for every examined $R_0$ value is the following in decreasing order: Preferential Attachment, Preferential Attachment mixed with ER, Erdős--Rényi graphs and random graphs generated by the configuration model. Attack rate for the Preferential attachment model is less sensitive to the change in $R_0$: Even for the case $R_0=1.0$ the overall attack rate is almost $0.3$, while the curve of infected individuals still has high kurtosis, with a maximum of $500$ individuals at the climax. From this aspect the models are very different: With bigger $R_0$ values, especially with $R_0=1.7$ overall attack rates are relatively close to each other, while the number of individuals in $I$ in Preferential attachment graphs is at least $2-3$ times bigger than in the other three models.
\\
The virus spread resulted in an escalated epidemic on every studied random graphs for every possible $R_0$ values, except for $R_0=[1.0,1.1]$ on Erdős--Rényi graphs and on random graphs from configuration model: For these parameters the number of infected individuals just gradually changed in time. In any other case the number of infected individuals in time is either a symmetrical curve, or a curve with a minimal positive skew for every age group obtained with the uniform vaccination strategy.

\subsection{Different vaccination strategies}
In this section we investigate how different strategies in vaccination can affect the attack rates. We study three very different strategies based on age groups or other properties of the graph. In each strategy $0.67\%$ of the population is vaccinated at each time step (sometimes exactly, sometimes only in expected value). After a $90$ days vaccination campaign $60\%$ of the population should be vaccinated from each age group (if it is possible). We still start our vaccination campaign at day 1, and we vaccinate individuals at most once irrespectively of the success of the vaccination.
\begin{itemize}
    \item Uniform strategy: This strategy does not distinguish age groups, every day we vaccinate randomly $0.67\%$ individuals of each age group.
    \item Contacts strategy: We prioritize age groups with bigger contact number, corresponding to denser parts of the graph (concerning the 5 groups). We vaccinate the second age group for 11 days, then the third age group for 26, first age group for 10 days, forth group for 29 days, and at last age group 5 with the smallest number of contacts for 15 days. This strategy turned out to be the best in the case without any graph structure \cite{1}.
    \item Degree strategy: With the exact knowledge of the underlying graph, we can not only favour the denser parts of the graph, but also specific nodes with the highest degree. Of course this strategy of vaccination is rather unrealistic. 
    At each time step we vaccinate exactly $67$ nodes with the highest degree still staying in $S$.
\end{itemize}

Firstly, we are curious whether the outcome of the process can be significantly different with different strategies. The best possible vaccination strategy could be different for various graph structures and also for different intensity of the disease. The other case is that there exists a global best strategy.

\begin{figure}[h!]
    \centering
    \includegraphics[height=6cm]{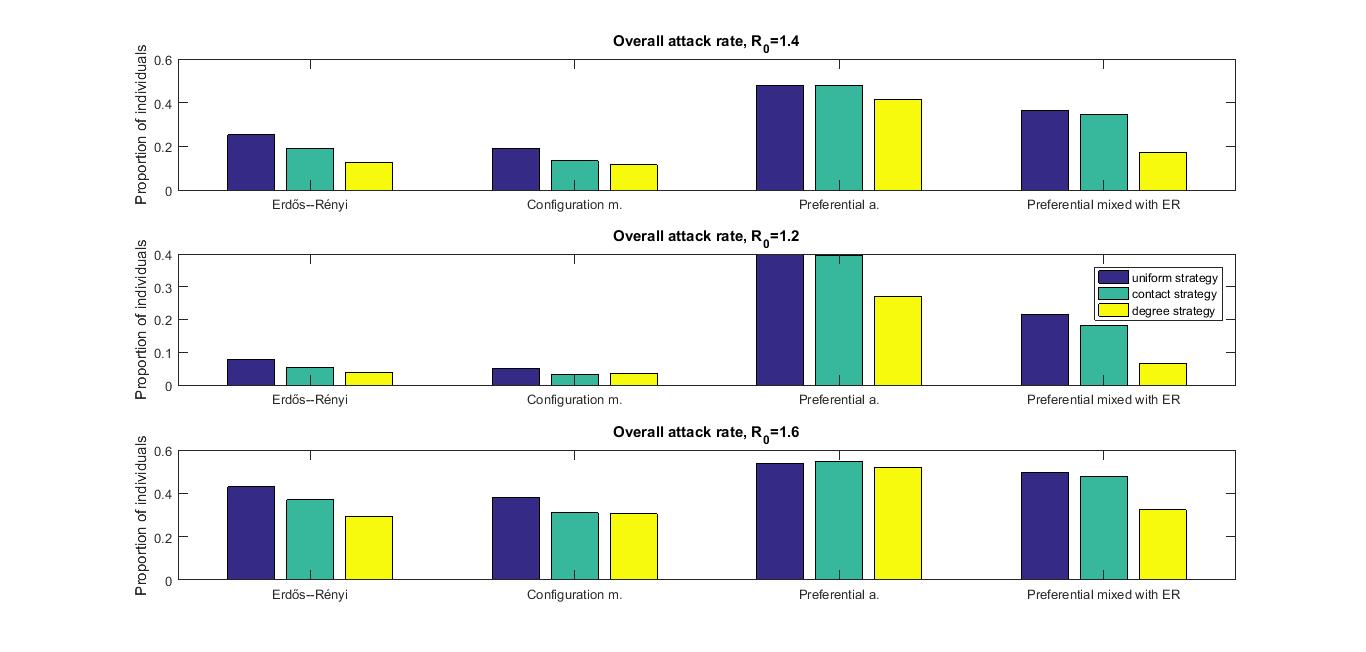}
    \caption{Overall attack rates with different vaccination strategies}
    \label{fig:strat}
\end{figure}

Figure \ref{fig:strat} shows that for every graph structure and for any reproduction number $R_0=1.2, 1.4, 1.6$,  vaccination by the degree of nodes is the best strategy. The difference between the effectiveness of the strategies are more outstanding in case of graph structures with (partial) preferential attachment properties, especially for smaller $R_0$ values. With a less severe virus, nodes with higher degrees vaccinated in the first steps of the process are more likely to survive the necessary 14 days for immunity without becoming infected, thus infection cannot spread at the numerous edges of these nodes. However, with bigger $R_0$ epidemic seems inevitable. On (partial) preferential attachment models attack rates with uniform vaccination or by contact strategies can be $1.5-2$ times bigger than with vaccination by degrees.
\\
Even for the configuration model we can see the difference in the outcomes, however vaccination by contact strategy gives close values to vaccination by degrees, since many nodes with high degrees coincide with nodes being in denser age groups. 
Unfortunately, in case of a real underlying network determining the exact number of edges of a node is rather challenging. They can be changing in time, especially with actions taken against the spread of the disease: closing schools, quarantine, shutting down public transportation. Even after calculating possible degrees of nodes, execution of vaccination could be problematic. \\
However, in conventional vaccination strategies, in the first days of the campaign, amongst others health care personnel is vaccinated which certainly makes sense, but can be also interpreted as nodes of the graph not only with high degree, but also with high probability to get infected.

\begin{figure}
    \centering
    \includegraphics[height=4cm]{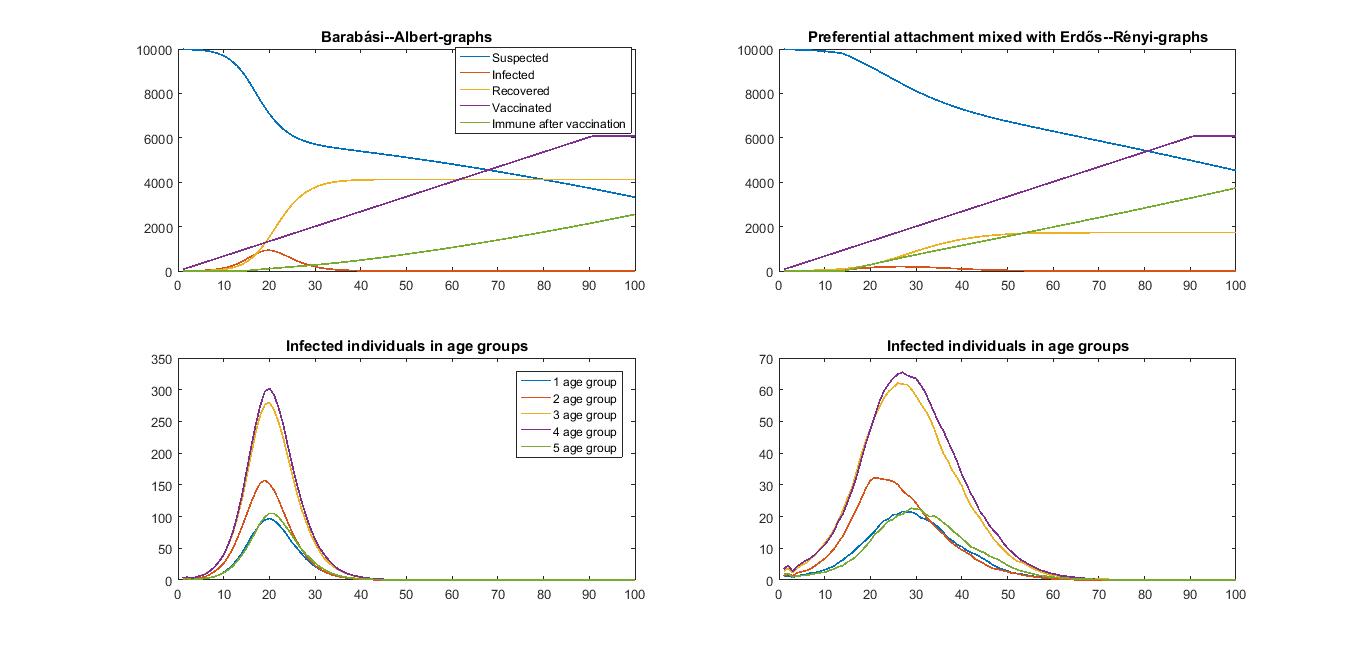}
    \includegraphics[height=4cm]{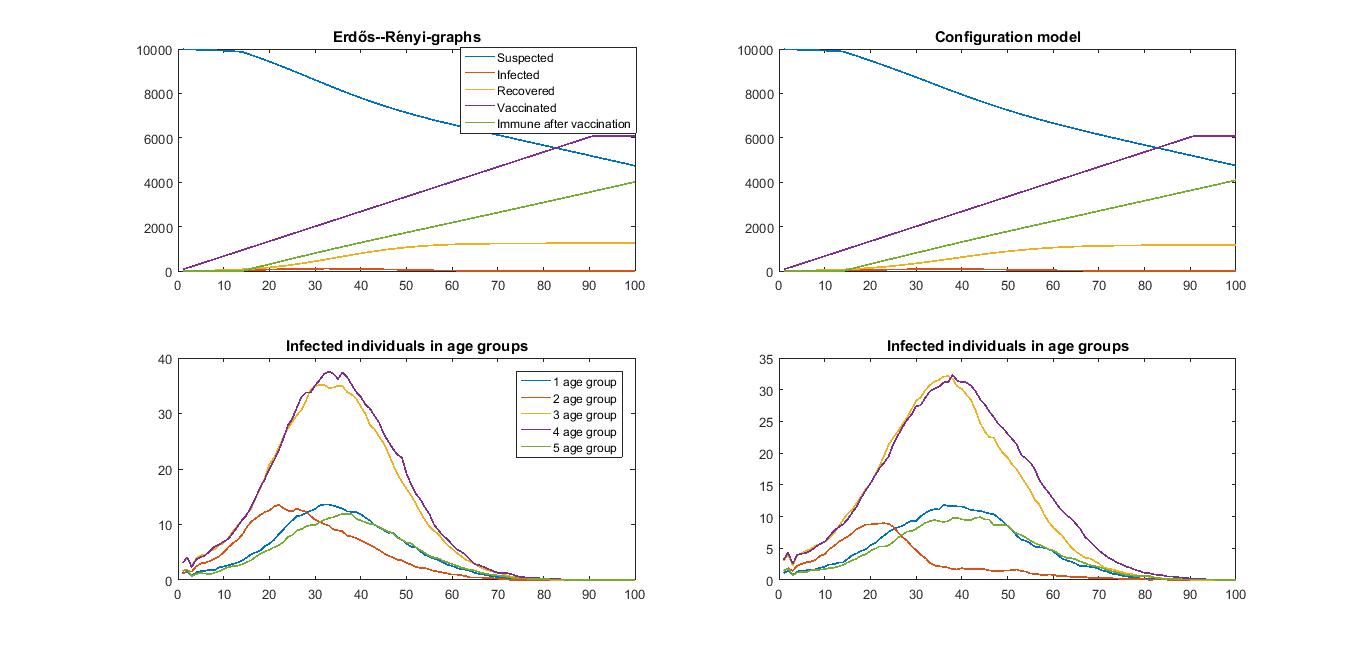}
    \caption{Vaccination by degrees, $R_0=1.4$}
    \label{fig:basicdeg}
\end{figure}

The effect of vaccination by degrees can be also noticed on the shape of infected individuals in age groups developing in time (see Figure \ref{fig:basicdeg}). Not only the magnitude decreased, but the vaccination also increased the skewness, especially for age group 2. Vaccination by contacts totally distorted the curve of age group 2, while the others did not changed much.

\subsection{Vaccination before infections}
We examine if vaccination before the outbreak of a virus (only a few, 5--10 days before) could influence the epidemic spread significantly. Delay in development of immunity after vaccination is one of the key factors of the model, thus pre-vaccination could counterweight this effect.

\begin{figure}[h!]
    \centering
    \includegraphics[height=5cm]{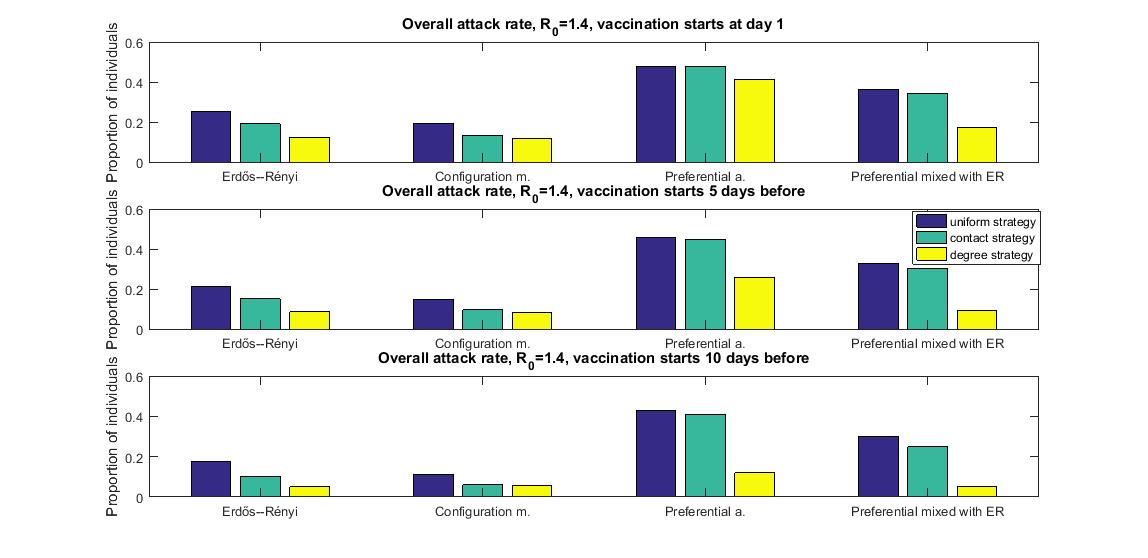}
    \caption{Attack rates with vaccination before 0, 5, 10 days of the outbreak}
    \label{fig:elozetes}
\end{figure}

As we can see on Figure \ref{fig:elozetes}, in some sense we got result similar to the behaviour of attack rates with different $R_0$ intensities of the epidemic. Graphs with (partial) preferential attachment structure reacted significantly to pre-vaccination only with vaccinating by degrees (with a maximum of 0.06 decrease in attach rates even for a 10-day period). For the configuration model results of vaccination by contacts and degrees behaved the same again. However, on Erdős--Rényi graphs and on the configuration model attack rates by uniform vaccination could almost decreased to half. For these structures, with any strategy, significant improvement can be achieved with a pre-existing vaccination.

\subsection{Edges with different weights}
Edges of the graph so far represented only the existence of a social contact, however relationships between the individuals could be of different quality. It is also a natural idea to make a connection between the type of the nodes (age groups of individuals) and the feature of the edge between them. For example, generally we can assume that children of age 0-9 (age group one) are more likely to catch or transmit a disease to any other individual regardless of age, since the nature of contacts with children are usually more intimate. So on the one hand, creating weights on the edges of the graph can strongly be in connection with the type of the given nodes. On the other hand regardless of age groups, individuals tend to have a few relationships considered more significant in the aspect of a virus spread (individuals sharing a household), while many social contacts are less relevant.

For the reasons above, we upgrade our random graphs with a weighting on the edges, taking into account the age groups of the individuals. Regardless of age, relationships divided into two types: close and distant. Only $20\%$ of the contacts of an individual can be close, transmission rates on these edges are much higher, while on distant edges they are reduced.

We examine a model in which age groups do not affect weights of the edges. We double the probabilities of transmitting the disease on edges representing close contacts, and decrease probabilities on other edges at a $0.75$ rate. In expected value the total $R_0$ of the disease has not changed. However, results on graphs can be different from the unweighted cases. 
With the basic scenario and in case of $R_0=1.4$ and $1.2$, for most graph differences in outcome are not significantly measurable $\mathcal{O}(0.001)$. We experience the biggest difference on Erdős--Rényi-graphs, however models with edge weights give bigger attack rates of only $0.01$. We get a less severe virus spread with edges only on the configuration model.

\newpage

\section{Voter model}

In this section we study the discretized voter model in which particles exchange opinions from time to time, in connection with relationships between them. We create a simplified process to be able to examine the outcome on larger graphs. Firstly, we examine this simplified process on Erdős--Rényi and Barabási--Albert graphs, then multiple type of nodes is introduced.
With a possible interpretation of different types of nodes in the graphs, we generalize the voter model. Later we examine the "influencer" model, in which our aim is, in opposition to the SEIR model, to spread one of the opinions. 

\subsection{Discretized voter model}

In the voter model an  undirected graph  $G(V,E)$ is given. The individuals are represented by the $n$ nodes of the graph, contacts between them are the edges. Initially each node has an opinion represented by a number in $\{0,1\}$: Each node independently chooses opinion 1 with probability $v$, and opinion 0 with $1-v$. Then individuals can change their opinion randomly in time, under the influence of other vertices. In case of epidemic spread (Section 2), infections can spread along edges, but vertices can pass the infection only to their neighbors. In the general voter model, interaction is possible between any pair of vertices.  However, the frequency  of the event that  vertex $x$ convinces vertex $y$ depends on the distance of $x$ and $y$, which we denote by $d(x,y)$. 

The process in continuous time can be modelled with a family of independent Poisson process. For each pair of vertices $(x, y)$ we have a Poisson process of  rate $q(x,y)$, which describes the  moments $x$ convincing $y$. The rate $q(x,y)$  increases as the distance $d(x,y)$ decreases. In this case, every time a vertex is influenced by another one, it changes its opinion immediately. 

In our discretized voter process, there are two phases at each time step. First, nodes try to share their opinions and influence each other, which is successful with probabilities depending on the distance of the two vertices. More precisely, vertices that are closer to each other have higher chance that their opinion "reaches" the other one.  Still, every vertex can "hear" different opinions from many other vertices. In the second phase, if a node $v$ receives the message of $m_0$ nodes with opinion $0$, and $m_1$ nodes with opinion $1$, then $v$ will represent opinion $0$ with probability $\frac{m_0}{m_0+m_1}$ during the next step, and $0$ otherwise. If a  node $v$ does not receive any opinions from others at a time step, then its opinion remains the same. This way, the order of influencing message in the first phase can be arbitrary, and it is also possible that two nodes exchange opinions.

Now we specify the probability that a vertex $x$ manages to share its opinion to vertex $y$ in the first phase. We transform  graph distances $d(x,y)$ into a matrix of transmission probabilities with choice $q(x,y)=e^{-c \cdotp d(x,y)}$, where $c$ is a constant. This is not a direct analogue of the continuous case, but it is still a natural choice of a decreasing function of $d$. (Usually we use $c=2$, however later we also investigate cases $c \in \{0.5,1,2,3\}$. Decreasing $c$ escalates the process.)

In the model above, on a graph on $n$  nodes, at every time step our algorithm consists of  $\mathcal{O}(n^2)$ steps, which can be problematic for bigger graphs if our aim is to make sample with $viter=100$ or $200$ iteration of the voter model (in the sequel, $viter$ denotes the number of steps of the voter model). However, with $c=2$ a node $x$ convinces vertices $y$ with $d(x,y)=3$ only with a probability of $e^{-6}=0,0025$. Thus we used the following simplified model: When we created a graph, we stored the list of edges and also calculated for each node the neighbours of distance 2. The simplified voter model spread opinions only on these reduced number of edges with the proper probabilities.
We were able to run the original discretized model only on graphs with $n=100$, while the simplified version can deal with $n=1000$ nodes. We made the assumption that neglecting those tiny probabilities cannot significantly change the outcome of the process. From now on we only model the simplified version of the process.

\subsection{Results on Erdős--Rényi and Barabási--Albert graphs}

Firstly we study the voter model on Erdős--Rényi$(n,p)$ and Barabási--Albert$(n,m)$ random graphs.
\begin{itemize}
    \item $ER(n,p)$: We create $n$ nodes, and connect every possible pair $x,y \in V$  independently with probability $p$.
    \item $BA(n,m)$: Initially we start with a graph $G_0$. At every time step we add a new  node $v$ to the graph and attach it exactly with $m$ edges to the old nodes with preferential attachment probabilities. Let $D$ denote the sum of degrees in the graph before adding the new node, then we attach an edge independently to $u$ with probability $\frac{d(u)}{D}$.
    \\ We generated graphs starting from $G_0=ER\big(50,\frac{m}{(50-1)}\big)$ graph of complying density. Multiple edges can be created by the algorithm, however loops cannot occur. Attachment probabilities are not updated during a time step. Multiple edges do matter in the voter model, since they somehow represent a stronger relationship between individuals: opinion on a $k$-multiple edge transmits with a $k$-times bigger probability.
\end{itemize}

\begin{figure}[h!]
    \centering
    \includegraphics[height=5cm]{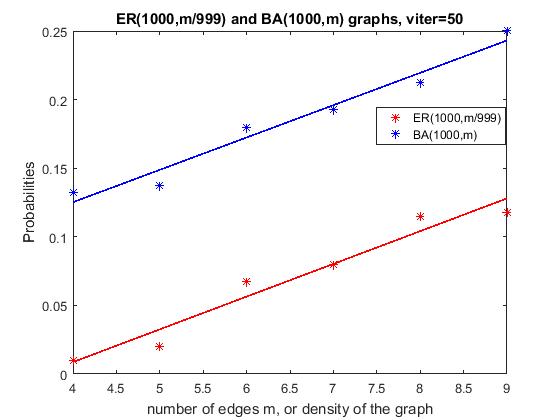}
    \caption{Probabilities of disappearing opinion 1 with variable graph density}
    \label{fig:voterdef}
\end{figure}

Firstly, we examine the voter model on graphs without any nodes of multiple types to understand the pure differences of the process resulting from the structure.  We compare graphs with the same density, $BA(1000,m)$ graphs with $m=\{4,5, \dots ,10 \}$ and $ER(1000,p)$, where $p\in [0.004,0.01]$. Initial probability of opinion 1 is set to $0.05$ in both graphs. We compare the probability of disappearing the opinion with $viter=50$ iteration of the voter model. We generated 10 different graphs from each structure and ran voter model on each 20 times with independent initial opinions. Altogether the results of 200 trials were averaged. Figure \ref{fig:voterdef} shows the results.

Before the phase transition of Erdős--Rényi graphs, that is, with  $p<\frac{\ln n}{n} \approx 0.007 $ with $n=1000$ nodes ($BA$ graphs of the same density are belonging to $m \leq 7$) the graph consists of several components with high probability. It is possible that nodes being in a tiny component of the graph, or even isolated nodes get opinion 1 initially, resulting in the co-existence of the two opinions. (In a small component of the graph within a few time steps any opinion can disappear easily, remaining the same for the rest of the process since no other nodes can influence them.) This could be one of the reasons why disappearing probabilities in $ER$ graphs are significantly less. The another reason is in connection with the next part: Since some nodes in $BA$ graphs can possess outstanding number of degrees, while most of them have only a few, the process can be influenced by the properties of nodes chosen to represent opinion 1 initially.
On the other hand, on $BA$ graphs not only the probability of the vanishing of  opinion 1 is higher, but it is also more likely to get extreme results. Proportions of opinion in $ER$ graphs are more stable.
We can generally say that the increase of the density of graphs escalates voter model, since in expected value more convictions happen at every time step, resulting in a more volatile proportion of opinions, and thus in higher probabilities of the disappearance of the underrepresented opinion.

\subsubsection{Different choices of $L_0$}
\label{diff L0}
As mentioned before, in this sequel we investigate extreme outcomes of the process caused by one of the most important properties of Barabási--Albert graphs.
Since nodes do not play a symmetrical role in Barabási--Albert graphs, fixing the proportion of nodes representing opinion 1 (we  usually use $v=0.05$, so 50 nodes represent opinion 1 in expected value), but changing the position of these nodes in the graph can lead to different results. We examined the following three ways of initial opinion setting:
\begin{itemize}
    \item randomly: Each individual chooses opinion 1 with probability $v$.
    \item "oldest nodes": We deterministically set the first 50 nodes of the graph to represent opinion 1. These nodes have usually the largest number of degrees, thus they play a crucial part in the process. Not only have they large degrees, but they are also very likely to be connected to each other (this is the densest part of the graph).
    \item "newest nodes": We deterministically set the last 50 nodes of the graph to represent opinion 1. These nodes usually have only $m$ edges, and they are not connected to each other with a high probability.
\end{itemize}

\begin{figure}[h!]
    \centering
    \includegraphics[height=6cm]{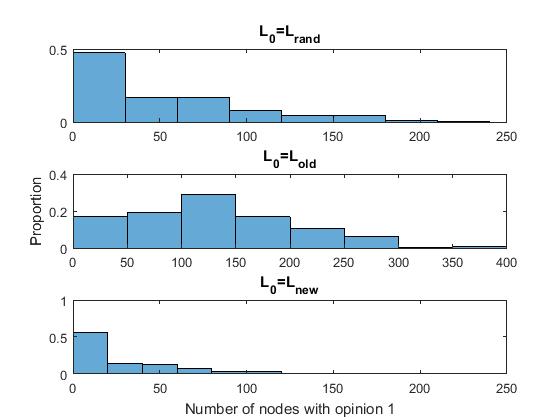}
    \caption{Distribution of nodes with opinion 1, different $L_0$}
    \label{fig:voterl0}
\end{figure}

The histogram on Figure \ref{fig:voterl0} shows the distribution of nodes with opinion 1 with the three different choice of $L_0$ vectors after $viter=50$ iterations of the voter model on $BA(1000,5)$ graphs. We experience differences in terms of probabilities of disappearing opinion 1: with random opinion distribution $11\%$, with $L_{new}$ almost one third of the cases resulted in extinction of opinion 1, while for $L_{old}$ this probability was negligible ($0.005\%$). Actually, for $L_{old}$ after only one iteration of the voter model it is impossible to see any structure in the distribution of individuals with opinion 1. Vector of opinions became totally random, but with a probability of $0.12$. Indeed only with one step of the voter model individuals with opinion 1 could double in number, however opinion 1 cannot take advantage of any special positions in the graph anymore.
All in all, giving a certain opinion to individuals who are more likely to be connected in the graph, reduces the probability of disappearing, since they can keep their opinion with a high probability, while with opinion 1 scattered across the graph (in case of $L_{new}$ as well as $L_{rand}$) with a dynamic parameter setting of $c$ number of individuals with opinion $1$ can reduce drastically even in a few time steps.

\subsection{Nodes with multiple types}
It is a natural idea to divide the nodes of a network into separate groups according to some aspect, where the properties of different groups can affect processes on the graph. There are various ways to classify nodes into different types. We examined a simple and an other widely used method. In the following section we only have nodes with two  types, however definitions still hold for multiple type cases. From now on, for purposes of discussion we only refer to the types as red and blue.

We consider two different ways to assign types to the nodes:
\begin{itemize}
    \item Each node independently of each other chooses to be red with probability $p_r$, and blue with $1-p_r$. (Here index $r$ corresponds to random.)
    \item Since preferential attachment graphs are  dynamic models, this enables another very natural and logical way of  choosing types: After a new node has connected to the graph with some edges, informally the node chooses its type with probabilities corresponding to the proportions among its neighbours' type (see also \cite{2, jordan, rosengren}). This way nodes with the same type tend to connect to each other with a higher probability, forming a "cluster" in the graph.
\end{itemize}

\begin{definition}
Preferential attachment graph with multiple types.
 
Initially we start with a $G_0$ graph, where nodes have a type of red or blue. At every time step we add a new  node $v$ to the graph and attach it with $m$ edges to the old nodes according to preferential attachment. Let $b$ denote the number of nodes with type blue attached to new node $v$. Then $v$ chooses to be blue with probability $p_b$, and red with probability $1-p_b$. The probability $p_b$ depends on $b$, as the index shows. We can define these probabilities in several ways in the function of $b$ number. The model is said to be linear if $p_b= \frac{b}{m}$ for every possible $b$ value, which is the natural choice.
In case of any other $p_b$ values the model is said to be nonlinear.
\end{definition}

We only examined linear models. According to $[2]$, a few properties in the initial  graph $G_0$ and initial types of nodes can determine the asymptotic behaviour of the proportion of types.

Let $G_n$ denote the graph when $n$ nodes have been added to the initial graph $G_0$. Let $A_n$ and $B_n$ denote the number of red and blue nodes in $G_n$. Then the following theorem holds for the asymptotic proportion of red ($a_n$) and blue ($b_n$)nodes, $a_n=\frac{A_n}{A_n+B_n}$, $b_n=\frac{B_n}{A_n+B_n}$. Let $X_n$ and respectively $Y_n$ denote the sum of the degrees of red and blue nodes in $G_n$.

\begin{Theorem}[\cite{2} Theorem 1.1, Linear model]
\label{tetel_multinodes}
Suppose that $p_b = \frac{b}{m}$ for all $0 \leq b \leq m$, and that $X_0, Y_0 \geq 0$. Then $a_n$ converges almost surely as $n \to \infty$. Furthermore, the limiting distribution of $a := lim_{n \to \infty}a_n$ has full
support on the interval $[0, 1]$, has no atoms, and depends only on
$X_0, Y_0$, and $m$.
\end{Theorem}

This property has great significance, since we would like to compare graphs with the same proportion of red and blue nodes. The theorem ensures us about the existence of such a limiting proportion. What is more, with the generation of Barabási--Albert graphs with multiple edges we can examine the speed of convergence. We set types of nodes in the initial graph $G_0$ in such a way that not necessarily half of the nodes will be blue, but approximately the sum degree of nodes with type blue will be the half of the whole sum of degrees. (Of course, in case of an initial Erdős--Rényi graph these will be the same in expected value. However we can get more stable proportion of types with the second method. In this case by stable we mean proportions can be closer to $\frac{1}{2}$.)

\subsection{Nodes with multiple types in the voter model}
In the voter model we can see nodes with multiple types, defined in the last section, with the following interpretation. Each node (individual) has two types according to two different aspects:
\begin{enumerate}
\item according to the ability to convince
    \begin{itemize}
    \item good reasoner type
    \end{itemize}
    \begin{itemize}
    \item bad reasoner type
    \end{itemize}
\item according to stability of opinion
    \begin{itemize}
    \item stable type
    \end{itemize}
    \begin{itemize}
    \item unstable type
    \end{itemize}
\end{enumerate}

So each node chooses a type from both of the aspects, and the choice of types according to different aspects are independent. (Since four combination of these is possible, we could say that each node chooses one type from the $4$ possible pairs.) During the voter model, interaction of nodes with different types influence the process in the following way: Complying with the names of the types, we expect that  good reasoner nodes could convince any nodes with a higher probability than bad reasoner nodes. Also any node should convince a node of unstable type with a higher probability than a node of a stable type. In a step of the voter model, when node $x$ influences a node $y$, the probability of success should only depend on node $x$'s ability to convince (good/ bad reasoner type) and node $y$'s stability of opinion. Instead of using a $c$ constant in the probability of success $q(x,y)=e^{-c \cdotp d(x,y)}$  of the voter model, value of $c$ will be different in case of different pairs of the types.

\begin{center}
\begin{tabular}{ | c | c | c |}
\hline
types & good reasoner & bad reasoner \\ 
\hline
stable & c(1) & - \\ 
\hline
unstable & c(2) & c(1) \\ 
\hline
\end{tabular}
\end{center}

We investigated the model with symmetric parameter set: The probability of a good reasoner node convincing a stable one is equal to the probability of a bad reasoner node convincing an unstable one. We also made the assumption that a bad reasoner node can convince a stable node with probability $0$. Voter model was examined with different $c(1) \geq c(2)$ set of parameters, and different possible choices of types in the graph.

\subsection{Erdős--Rényi and Barabási--Albert graphs with multiple type nodes}

In this sequel we examine a special case of voter model with multiple type nodes, in which the aim is to spread an initially underrepresented opinion. This problem might be related to finding good marketing strategies on online social networks, when "opinion" might be about a commercial product or a certain political convinction. 

\begin{figure}[h!]
    \centering
    \includegraphics[height=5cm]{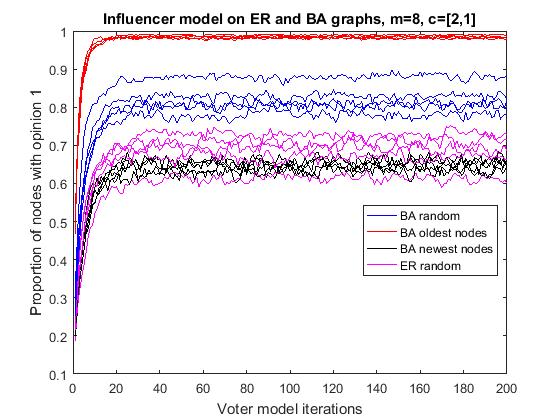}
    \caption{Proportion of opinion 1 with different  choices of $L_0$ in $ER$ and $BA$ graphs}
    \label{fig:influencer_erba}
\end{figure}

We investigate the following "influencer" model: Types of a node according to the different aspects is not independent, nor is the $L_0$ vector of initial opinions. The nodes of the graph are divided into two groups, influencers and non-influencers. Influencers  usually form a smaller population; they represent opinion $1$, which we want to spread across the graph. They are good reasoners, and also stable, while non-influencers have bad reasoner type according to the ability to convince, while they can be stable as well as unstable. According to definitions of $c$ values, it is impossible for a bad reasoner node to convince a stable one, resulting influencers representing opinion 1 for the whole process.
\\
Firstly, we study a case in which nodes of a $BA$ graph get a type randomly or deterministically, not according to preferential attachment. We study the equivalent of the case in subsection \ref{diff L0} with multiple type nodes. In each graph $10\%$ of individuals (100 nodes) are influencers. In $BA$ graphs influencers are situated randomly, on the "oldest nodes" or on the "newest nodes" of the graph. In $ER$ graphs influencers are situated randomly (however, since the role of nodes is symmetric, they can be situated anywhere, with no difference in the outcome).
We would like to examine the differences in opinion spread. We are also interested whether it is possible to convince all the nodes of the graph to opinion $1$, and in case it is, we calculate the average time needed to do so.
\\
We observed differences in the outcome for 100 runs (on 5 different random graphs), with $c=[2,1]$ and $m=8$ parameter set (see Figure \ref{fig:influencer_erba}). Since both Erdős--Rényi and Barabási--Albert graphs have small-world network properties, opinion 1 spreading to more than half of the nodes (only after a couple of iterations of the voter model) even in the worst cases is not a surprise. In fact, proportion of nodes with opinion 1 is stable after $viter=20$ steps of the voter model, with a mean of 0.82; 0.98; 0.65; and 0.66 corresponding to the different positions of influencers and graphs. For the two cases of randomly chosen initial  opinions $L_0$, variance of the proportions is significantly greater than in the deterministic cases. However, this difference in variances seems to derive only from the initial choices of influencers: After 20 iterations not only the mean of the trajectories is stable, but each trajectory itself. The magnitude of proportion of opinion $1$ after 20 time steps depends on the degrees of randomly chosen influencers.
What is more surprising is that even on $BA$ graphs with choosing the "oldest" nodes to be influencers the coexistence property of trajectories still holds. Even after $200$ steps of the voter model, opinion $0$ did not disappear  in any of the cases (from 100 runs), despite low $(\approx 0.02)$  proportion.

Secondly, we examined the influencer model with the nodes of $BA$ graphs getting types according to preferential attachment dynamics. However, according to Theorem \ref{tetel_multinodes}, although the proportion of nodes with different types converges, setting exactly a given proportion can be rather challenging. (We wanted to exclude cases in which the proportion of one of the types is negligibly small.) For these reasons, we created $ER$ and $BA$ graphs with multiple nodes, where the proportion of good reasoners is $\frac{1}{2}$ and according to preferential attachment (in $BA$ graphs), we set these nodes to be the influencers (they are stable, while non-influencer individuals can be stable and unstable with probability $\frac{1}{2}$). So in expected value half of the nodes are influencers, but in case of $BA$ graphs we can experience greater deviance (in expected value half of the nodes have good reasoner type according to the ability to convince, and $\frac{3}{4}$ of the nodes have stable type). For $ER$ graphs the only meaningful possibility to create types is the random choice, but the same proportions also hold.
\\

With half of the population being influencers, opinion $0$ can disappear even with $c=[3,1.5]$ and $c=[4,2]$ less dynamic parameters. Hence in some cases it is possible to talk about the average time of disappearance of opinion $0$.
We examined 100 runs on 10 different random graphs from both structures: In this case after $viter=200$ steps of the voter model, there is only an insignificant difference between the mean proportion of opinion $1$ on the two graph structures (0.9948 and 0.9965, proportions in $BA$ graphs are still a bit greater). However, in terms of disappearing opinions results are rather different. On $ER$ graphs in none of the cases could opinion $0$ disappear, while on $BA$ graphs it strongly depended on the exact initial proportion of influencers in the graph: On the same graph (and hence with the same proportion of influencers) opinion $0$ disappears either within the first $50$ iterations of the voter model, or holds a high proportion of opinion $1$, yet it will never be able to reach the limit. This main difference resulted from the fact that we can not exactly set the proportion of types in $BA$ graphs, thus the co-existence of opinions is rather sensitive to changes in the number of influencers in $BA$ graphs. (In $ER$ graphs only $20\%$ of the examined runs resulted in disappearance of opinion $0$, even with $600$ influencers.)

\subsection{Random graph on a plane}

In this section we examine the voter model on a random graph which has a geometric structure on the plane. Since the graph model is not dynamic, nodes can only choose their type randomly (or according to some deterministic strategy related to the position of nodes in the plane). However firstly we study the model without multiple types, with constant $c=0.5,0.1$.

\begin{definition}[Random Points on the square]
$RP(n)$ is a random graph on $n$ nodes. Each node has a position on the plane. Coordinates $(x,y)$ of node $v$ are generated randomly and independently from the interval  $[0,100]$. The nodes create a complete weighted graph, where the length of an edge between nodes $v$ and $u$ is  the classical Euclidean distance of their positions, denoted by $d(v,u))$.
\end{definition}

Since the voter model is rather time-consuming, and even in case of parameter $c=0.5$ the probability of conviction $q(x,y)=e^{-c \cdotp d(x,y)}$  for $d(x,y)=10$ is $\approx 0.0067$. Thus we create a reduced graph from $RP(n)$ by erasing edges in case of $d(x,y)>10$. We can assume that results on the reduced graph can approximate the outcome on the original one, since transmission of opinions on those edges are negligible. The average degree in the reduced graph is still $27.85$. Modifying the voter model to spread opinions only on these edges makes the algorithm less robust and manageable to run the process on graphs with many ($n=1000$) nodes.

\subsubsection{Different positions for $L_0$}
Firstly, we would like to understand the behaviour of the process without multiple types of the graph. In this section we make an advantage of the geometric structure of the graph, and examine different deterministic and random choices for initial opinions of $L_0$. We study how these alternative options can influence the outcome (the probability of the disappearance of an opinion, expected time needed for extinction). Another interesting question is whether after a given number of iterations $t$ of the voter model we can still observe any nice shape of the situation of opinions. 
In both of the following four choices for initial opinions in expected value $10\%$ of the individuals are given opinion $1$, the rest of them represent opinion $0$.
\begin{itemize}
    \item Random: Each individual represents independently opinion $1$ with probability $0.1$
    \item Corner: Individuals with opinion $1$ are clustered into one of the corners of the square $[0,100] \times [0,100]$, into square $[0,31.62] \times [0, 31.62]$.
    \item Outside: In this case individuals representing opinion $1$ are scattered to the "outside" region of the graph, in particular, outside square $[2.57, 97.43] \times [2.57, 97.43]$.
    \item Inside: Individuals clustered again, but this time to the very centre of the graph, inside square $[34.19, 65.81] \times [34.19, 65.81]$.
\end{itemize}

\begin{figure}[h!]
    \centering
    \includegraphics[height=4cm]{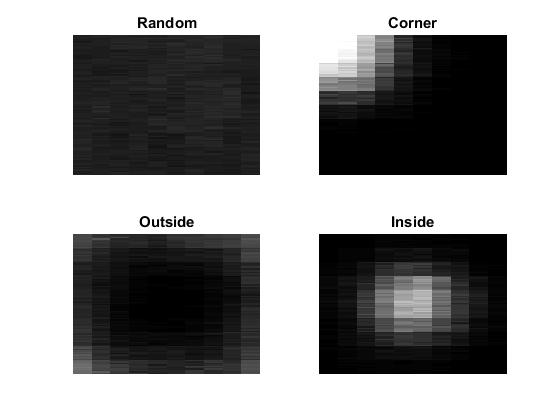}
    \caption{Positions of opinion 1 in $RP(1000)$ graphs for different $L_0$ vectors of initial opinion}
    \label{fig:rp1}
\end{figure}

The discretized voter model with different  initial opinion vectors $L_0$ was performed on 400 different graphs for $viter=100$ steps. With $n=1000$ number of nodes, $c=0.5$ and only $10\%$ of population representing opinion 1, opinion $1$ disappeared only in a few (negligible) cases   for any examined $L_0$. We can say, without any doubt according to picture \ref{fig:rp1}, that after 100 time steps the deterministic position of initial opinions is still recognizable (even after $viter=200$ steps).

\begin{figure}[h!]
    \centering
    \includegraphics[height=4cm]{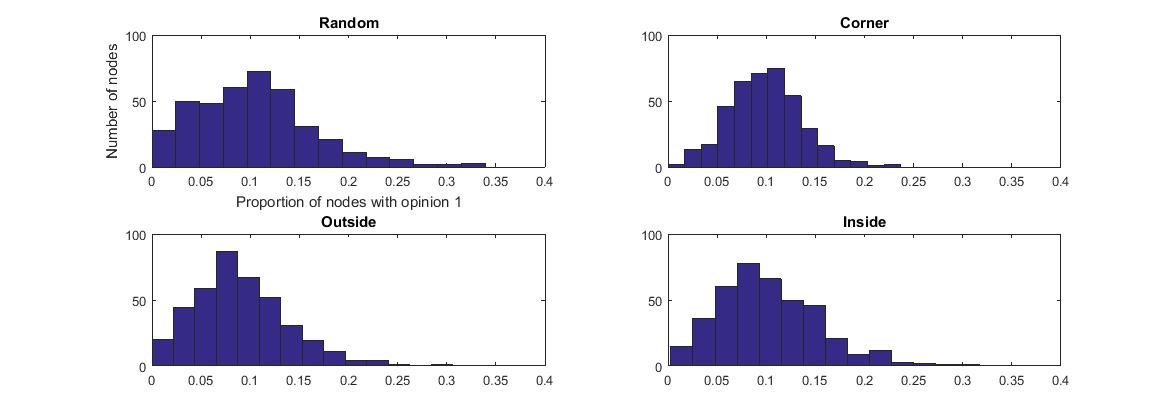}
    \caption{Proportion of opinion 1 in $RP(1000)$ graphs for different $L_0$ vectors of initial opinion}
    \label{fig:rp2}
\end{figure}

The histogram of proportion of individuals with opinion 1 on Figure \ref{fig:rp2} is more representative than only the number of cases with disappearing opinions. 

We can generally state that with clustering individuals with the same opinion in a group, makes proportion of opinions more stable in the process: From the different runs we observed that proportion of opinions (from initial $0.1$) stayed between $[0.8, 1.2]$ with a probability of more than $0.4$ in case of opinion 1 situated in a corner of the graph, while in any other cases this was significantly lower (less than $0.3$). With this placement of opinion 1, average distances within individuals with opinion 1 was the smallest, however average distances between different groups of opinions was the largest among the examined cases, resulting in moderate change of opinions.
\\
Number of individuals representing 1 decreased below 50 only with probability of $0.08$, while with placing opinion 1 in the center this probability is $0.1325$. Opinion 1 is the most likely to disappear (with probability $0.195$), or reduce to an insignificant amount with random placement of opinion 1. However, inverse extreme cases are also more likely to occur, since proportions of opinion 1 exceeding $0.2$ is outstandingly high with this scenario. Moreover, despite the high probability of extinction, in expected value we get the highest proportion of opinion 1 after $viter=100$ iteration of the voter model with random initial configuration.

\subsubsection{Influencer model on random graphs on a plane}

We also examined random graphs on a plane with a random or deterministic type choice of the nodes corresponding to two different aspect as before. We set the type pairs to form an influencer model defined before. Due to the fact that average distances in this random graph model are significantly larger than in $ER$ and $BA$ graphs of small-world property, small proportion of influencers in most of the cases (different position of influencers) can only reach a limited proportion of nodes. (With some position of influencers this limit is rather low, and can be reached within a few iteration of the voter model: Influencers gathered in a corner of the graphs the limit is below $15\%$ of the individuals, while the average proportion of opinion 1 is only $0.12$.)
\\ In most of the cases neither can help the problem the setting of all non-influencer individual to unstable type. Even with random influencer position the calculation of average time needed to convince all nodes of the graphs is challenging due to its time cost. With the increase of influencers in number to $300$, in half of the runs was able to reach opinion 1 all nodes of the graph within $400$ time steps. However, sometimes only in a relatively small number of iterations, suggesting that exact position on the plane of randomly chosen individuals do effect the process significantly.


\begin{thebibliography}{99}
\bibitem{2} Tonći Antunović, Elchanan Mossel, Miklós Z. Rácz, \emph{Coexistence in preferential attachment networks}, Combin. Probab. Comput. 25 (6) (2016), 797--822.
\bibitem{basu} Riddhipratim Basu, Allan Sly, \emph{Evolving voter model on dense random graphs},     Ann. Appl. Probab. 27 (2) (2017), 1235--1288.
\bibitem{berger}     Noam Berger Christian Borgs Jennifer Chayes Amin Saberi, \emph{On the spread of viruses on the internet}, Proceedings of the 16th ACM-SIAM Symposium on Discrete Algorithm (SODA), 2005.
\bibitem{bhansali} Rinni Bhansali, Laura P. Schaposnik, \emph{A trust model for spreading gossip in social networks}. Preprint. arXiv:1905.11204
\bibitem{4} Béla Bollobás, \emph{Random Graphs}. Second Edition, Cambridge University Press, 2001.
\bibitem{britton} Tom Britton, \emph{On critical vaccination coverage in multitype epidemics}, J. Appl. Probab. 35 (1998), 1003--1006.
\bibitem{britton2} Tom Britton, Svante Janson, Anders Martin-Löf, \emph{Graphs with specified degree distributions, simple epidemics, and local vaccination strategies}, Adv. Appl. Probab. 39(4) (2007), 922-–948. 
\bibitem{carro} Adrián Carro, Raúl Toral, Maxi San Miguel, \emph{The noisy voter model on complex networks}, Scientific Reports 6, 24775 (2016).
\bibitem{durrett} Rick Durrett, Claudia Neuhauser, \emph{Coexistence results for some competition models}, Ann. Probab. 7 (1), (1997), 10--45.
\bibitem{durrett2} Rick Durrett, \emph{Random graph dynamics}, Cambridge University Press, Cambridge, 2006.
\bibitem{fransson} Carolina Fransson, Pieter Trapman, \emph{SIR epidemics and vaccination on random graphs with clustering}, J Math Biol.  78 (7) (2019) 2369–-2398.
\bibitem{hofstad} Remco van der Hofstad, \emph{Random graphs and complex networks}, Cambridge University Press, Cambridge, 2016.
\bibitem{jordan} Jonathan Jordan, \emph{Preferential attachment graphs with co-existing types of different fitnesses}, J. Appl. Probab. 55 (4) (2018), 1211--1227.
\bibitem{1} Diána H. Knipl, Gergely Röst
\emph{Modelling the strategies for age specific vaccination scheduling during influenza pandemic outbreaks}, Math. Biosci. Eng. 8 (1) (2011), 123--139.
\bibitem{simon} István Z. Kiss, Joel Miller, Péter Simon, \emph{Mathematics of Epidemics on Networks}, Springer, 2017. 
\bibitem{3} Thomas M. Liggett, \emph{Interacting Particle Systems}, Springer, New York, 1985.
\bibitem{rosengren} Sebastian Rosengren, \emph{A Multi-type preferential attachment tree}, Internet Math. September 07, 2018. https://doi.org/10.24166/im.05.2018.
\bibitem{5} Shi Zhao, Jinjun Ran, Salihu S Musa, Guangpu Yang, Yijun Lou, Daozhou Gao, Lin Yang, Daihai He, \emph{Preliminary estimation of the basic reproduction number of novel coronavirus (2019-nCoV) in China, from 2019 to 2020: A data-driven analysis in the early phase of the outbreak}, International Journal of Infectious Diseases. In press. https://doi.org/10.1016/j.ijid.2020.01.050


\end{thebibliography}
\end{document}